\documentclass{PoS}

\title{Transverse (Spin) Structure of Hadrons}

\ShortTitle{Transverse (Spin) Structure of Hadrons}

\author{\speaker{Matthias Burkardt}\\
        New Mexico State University\\
        E-mail: \email{burkardt@nmsu.edu}}

\abstract{ Parton distributions in impact parameter space, which
are obtained by Fourier transforming GPDs, exhibit
a significant deviation from axial symmetry when the
target and/or quark are transversely polarized. 
Connections between this deformation and transverse single-spin
asymmetries as well as with quark-gluon correlations 
are discussed. The sign of transverse deformation of impact parameter dependent parton distributions in a transversely polarized target can be related to the sign of the contribution from that quark flavor to
the nucleon anomalous magnetic moment. Therefore, the signs of the Sivers function for $u$ and $d$ quarks, as well as the signs
of quark-gluon correlations embodied in the polarized structure function $g_2$ can be understood in terms of the proton and neutron
anomalous magnetic moments.
}

\FullConference{Light Cone 2010 - LC2010\\
		June 14-18, 2010\\
		Valencia, Spain}

\newcommand{\be}{\begin{eqnarray}}
\newcommand{\ee}{\end{eqnarray}}
\newcommand{\bea}{\begin{eqnarray}}
\newcommand{\eea}{\end{eqnarray}}
\def\xbj{x_{\mbox{\tiny B}}}

\begin{document}

\section{Impact Parameter Dependent Parton Distributions}
\label{sec:IPDs}
Generalized Parton Distributions (GPDs) can be obtained from the
same light-cone wave function overlap integrals that yield
form factors, except that the momentum fraction $x$
of the active quark is not integrated over, i.e. GPDs can be
understood as an $x$ decomposition of form factors.
The 2-dimensional Fourier transform of the GPD 
$H_q(x,0,t)$ yields the  
distribution $q(x,{\bf b}_\perp)$ of 
unpolarized quarks and target, in 
impact parameter space \cite{mb1} 
\be
q(x,{\bf b}_\perp)= 
 \int \!\!\frac{d^2\Delta_\perp}{(2\pi)^2}  
H_q(x,0,\!-\Delta_\perp^2) \,e^{-i{\bf b_\perp} \cdot
\Delta_\perp}, \label{eq:GPD}
\ee 
with $\Delta_\perp = {\bf p}_\perp^\prime -{\bf p}_\perp$.
For a transversely polarized target (e.g. polarized in the
$+\hat{x}$-direction) the impact parameter dependent
PDF $q_{+\!\hat{x}}(x,{\bf b}_\perp)$ is
no longer axially symmetric and
the transverse deformation is described
by the gradient of the Fourier transform of the GPD $E_q(x,0,t)$
\cite{IJMPA}
\bea
q_{+\!\hat{x}}(x,\!{\bf b_\perp}) 
&=& q(x,\!{\bf b_\perp})
- 
\frac{1}{2M} \frac{\partial}{\partial {b_y}} \!\int \!
\frac{d^2\Delta_\perp}{(2\pi)^2}
E_q(x,0,\!-\Delta_\perp^2)\,
e^{-i{\bf b}_\perp\cdot\Delta_\perp}
\label{eq:deform}
\eea
$E_q(x,0,t)$ and hence the details of this deformation are not 
very well known, but its $x$-integral, the Pauli form factor
$F_2$, is. Eq. (\ref{eq:deform})  allows to
relate the average transverse deformation 
\be
d^q_y \equiv \int\!\! dx\!\int \!\!d^2{\bf b}_\perp
q(x,{\bf b}_\perp ) b_y
=\frac{1}{2M} 
\int\!\! dx E_q(x,0,0) = \frac{\kappa_{q}^p}{2M}  
\ee
to 
\begin{figure}
\unitlength1.cm
\begin{picture}(10,5)(2.7,12.7)
\includegraphics{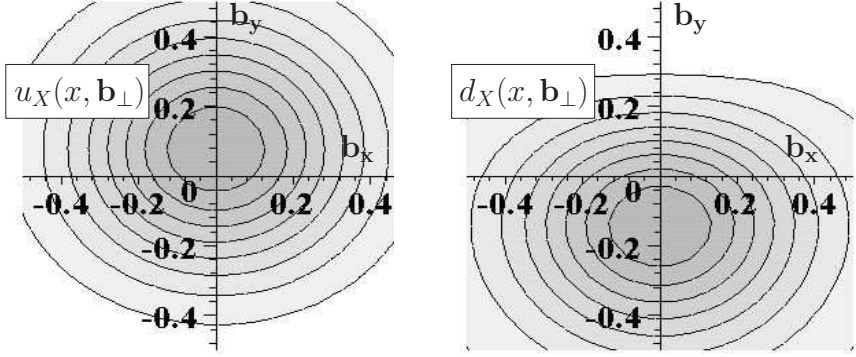}
\end{picture}
\caption{Distribution of the $j^+$ density for
$u$ and $d$ quarks in the
$\perp$ plane ($x_{Bj}=0.3$) for a proton 
polarized
in the $x$ direction in the model from
Ref. \cite{IJMPA}.
For other values of $x$ the distortion looks similar. The signs of
the distortion are determined by the signs of the contribution
from each quark flavor to the proton anomalous magnetic moment.
}
\label{fig:distort}
\end{figure}  
the contribution from the
corresponding quark flavor to the anomalous magnetic moment\\
$\kappa_u^p=2\kappa_p +\kappa_n= 2*1.793 - 1.913
= 1.673$ and
$\kappa_d^p=2\kappa_n+\kappa_p= 2*(-1.913) + 1.793
= -2.033$. Since $\frac{1}{2M}\approx 0.1\,fm$ this implies a very
significant deformation
$|d_q^y|={\cal O}(0.2\,fm)$ for both $u$ and $d$ quarks and
in opposite directions.

For example, $u$ quarks in a proton contribute with a positive 
anomalous magnetic moment and $d$ quarks (after factoring out the 
negative $d$ quark charge) with a negative value.
Eq. (\ref{eq:deform}) thus implies that for a nucleon
target polarized in the $+\hat{x}$ direction, the leading twist
distribution of $u$ quarks is shifted in the $+\hat{y}$ direction
while that of $d$ quarks is shifted in the $-\hat{y}$ direction
(Fig. \ref{fig:distort}).
This has important implications for the sign of transverse
single-spin asymmetries (SSAs).

\section{Transverse Single-Spin Asymmetries}

In a target that is polarized transversely ({\it e.g.} vertically), 
the quarks in the target 
can exhibit a (left/right) asymmetry of the distribution 
$f_{q/p^\uparrow}(\xbj,{\bf k}_T)$ in their transverse 
momentum ${\bf k}_T$ \cite{sivers,trento}
\be
f_{q/p^\uparrow}(\xbj,{\bf k}_T) = f_1^q(\xbj,k_T^2)
-f_{1T}^{\perp q}(\xbj,k_T^2) \frac{ ({\bf {\hat P}}
\times {\bf k}_T)\cdot {\bf S}}{M},
\label{eq:sivers}
\ee
where ${\bf S}$ is the spin of the target nucleon and
${\bf {\hat P}}$ is a unit vector opposite to the direction of the
virtual photon momentum. The fact that such a term
may be present in (\ref{eq:sivers}) is known as the Sivers effect
and the function $f_{1T}^{\perp q}(\xbj,k_T^2)$
is known as the Sivers function.
The latter vanishes in a naive parton 
picture since $({\bf {\hat P}} \times {\bf k}_T)\cdot {\bf S}$ 
is odd under naive time reversal (a property known as naive-T-odd), 
where one merely reverses
the direction of all momenta and spins without interchanging the
initial and final states. 
The significant distortion of parton distributions in impact 
parameter space (Fig. \ref{fig:distort})
provides a natural mechanism for a Sivers effect.
In semi-inclusive DIS, when the 
virtual photon strikes a $u$ quark in a $\perp$ polarized proton,
the $u$ quark distribution is enhanced on the left side of the target
(for a proton with spin pointing up when viewed from the virtual 
photon perspective). 
\begin{figure}
\unitlength1.cm
\begin{picture}(10,2.3)(3.,19.2)
\includegraphics{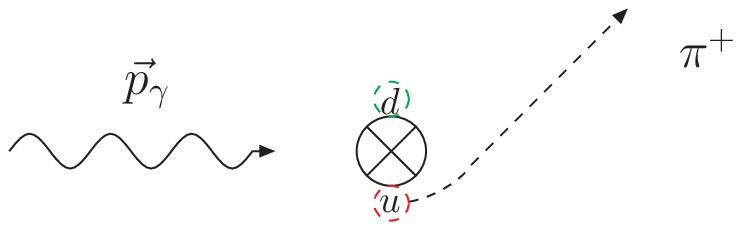}
\end{picture}
\caption{The transverse distortion of the parton cloud for a proton
that is polarized into the plane, in combination with attractive
FSI, gives rise to a Sivers effect for $u$ ($d$) quarks with a
$\perp$ momentum that is on the average up (down).}
\label{fig:deflect}
\end{figure}
Although in general the final state 
interaction (FSI) is very complicated, we expect it to be on average attractive thus translating a position space
distortion to the left into a momentum space asymmetry to the right
and vice versa (Fig. \ref{fig:deflect}) \cite{mb:SSA}.
Since this picture is very intuitive, a few words
of caution are in order. First of all, such a reasoning is strictly 
valid only in mean field models for the FSI as well as in simple
spectator models \cite{spectator}. 
Furthermore, even in such mean field or spectator models
there is in general
no one-to-one correspondence between quark distributions
in impact parameter space and unintegrated parton densities
(e.g. Sivers function) (for a recent overview, see Ref.
\cite{Metz}). While both are connected by an overarching Wigner
distribution \cite{wigner}, 
they are not Fourier transforms of each other.
Nevertheless, since the primordial momentum distribution of the quarks
(without FSI) must be symmetric, we find a qualitative connection
between the primordial position space asymmetry and the
momentum space asymmetry due to the FSI.
%
Another issue concerns the $x$-dependence of the Sivers function.
The $x$-dependence of the position space asymmetry is described
by the GPD $E(x,0,-{\Delta}_\perp^2)$. Therefore, within the above
mechanism, the $x$ dependence of the Sivers function should be
related to 
that of $E(x,0,-{\Delta}_\perp^2)$.
However, the $x$ dependence of $E$ is not known yet and we only
know the Pauli form factor $F_2=\int {\rm d}x E$. Nevertheless, 
if one makes
the additional assumption that $E$ does not fluctuate as a function 
of $x$ then the contribution from each quark flavor $q$ to the
anomalous magnetic moment $\kappa$ determines the sign of 
$E^q(x,0,0)$
and hence of the Sivers function. With these assumptions,
as well as the very plausible assumption that the FSI is on average
attractive, 
one finds that $f_{1T}^{\perp u}<0$, while 
$f_{1T}^{\perp d}>0$. Both signs have been confirmed by a flavor
analysis based on pions produced in a SIDIS experiment 
by the {\sc Hermes} collaboration \cite{hermes} and are
consistent with a vanishing isoscalar Sivers function observed
by {\sc Compass} \cite{compass}.

\section{Transverse Force on Quarks in DIS}
The chirally-even spin-dependent twist-3 parton distribution 
$g_2(x)=g_T(x)-g_1(x)$ is defined as
\begin{eqnarray}
& &\int \frac{d\lambda}{2\pi}e^{i\lambda x}
\langle PS|\bar{\psi}(0)\gamma^\mu\gamma_5\psi(\lambda n)
|_{Q^2}|PS\rangle
\nonumber\\
& &\qquad=
2\left[g_1(x,Q^2)p^\mu (S\cdot n) 
+ g_T(x,Q^2)S_\perp^\mu +M^2 g_3(x,Q^2)n^\mu (S\cdot n)
\right].
\nonumber
\end{eqnarray}
Neglecting $m_q$, one finds $g_2(x)=g_2^{WW}(x)+\bar{g}_2(x)$, with
$g_2^{WW}(x)=-g_1(x)+\int_x^1 \frac{dy}{y}g_1(y)$ \cite{WW},
where
$\bar{g}_2(x)$ involves quark-gluon correlations, e.g.
\cite{Shuryak,Jaffe}
\be
\int dx x^2 \bar{g}_2(x)= \frac{d_2}{3}
\ee
with
\be
4M {P^+}P^+ S^x d_2=
g\left\langle P,S \left|\bar{q}(0)G^{+y}(0)\gamma^+q(0) 
\right|P,S\right\rangle .
\label{eq:twist3}
\ee
At low $Q^2$, $g_2$ has the physical interpretation of a spin 
polarizability, which is why the matrix elements (note that
$\sqrt{2}G^{+y}=B^x-E^y$) 
\be
\chi_E 2M^2 {\vec S} = \left\langle P,S\right|
q^\dagger {\vec \alpha} \times g {\vec E} q \left| P,S\right\rangle
\quad\quad
\chi_B 2M^2 {\vec S} = \left\langle P,S\right|
q^\dagger g {\vec B} q \left| P,S\right\rangle
\ee
are sometimes called spin polarizabilities or color electric and 
magnetic polarizabilities \cite{Ji}. In the following we will 
discuss that at
high $Q^2$ a better interpretation for these matrix elements is that
of an average `color Lorentz force' \cite{mb:g2}.

To see this we express the $\hat{y}$-component of the Lorentz force 
acting on 
a particle with charge $g$ that is moving with (nearly) 
the speed of light
${\vec v}=(0,0,-1)$ along the $-\hat{z}$ direction in terms of 
light-cone variables, yielding
\be
g\left[ {\vec E} + {\vec v}\times
{\vec B}\right]^y = g\left(E^y + B^x\right) = 
g\sqrt{2}G^{y+},
\ee
which coincides with the component that appears in the twist-3
correlator above (\ref{eq:twist3}).
Thus Eq. (\ref{eq:twist3}) represents the (twist 2) 
quark density correlated with the transverse color-Lorentz force
that a quark would experience at that position if it moves with
the velocity of light in the $-\hat{z}$ direction --- which is 
exactly what the struck quark does after it has absorbed the virtual 
photon in a DIS experiment in the Bjorken limit. Therefore the 
correct semi-classical interpretation of Eq. (\ref{eq:twist3}) is
that of an average\footnote{The average is meant as an ensemble average since the forward
matrix element in plane wave states automatically provides an average
over the nucleon volume.} 
transverse force
\be
\label{eq:QS3}
F^y(0)&\equiv& - \frac{\sqrt{2}}{2P^+}
\left\langle P,S \right|\bar{q}(0) G^{+y}(0)
\gamma^+q(0) \left|P,S\right\rangle\\
&=& -2\sqrt{2} MP^+S^xd_2
= -2M^2d_2
\nonumber
\ee
acting on the active quark
in the instant right after\footnote{`Right after', since the 
quark-gluon correlator in (\ref{eq:QS3}) is local!} 
it has been struck by the virtual photon.

Although the identification of 
$\langle p | \bar{q}\gamma^+ G^{+y}q|p\rangle$ as an average 
color Lorentz
force due to the final state interactions (\ref{eq:QS3})
may be intuitively evident from the above discussion, it is 
also instructive to 
provide a more formal justification. For this purpose, we consider
the time dependence
of the transverse momentum of the `good'
component of the quark fields (the component relevant for 
DIS in the Bjorken limit)
${q}_{+} \equiv
\frac{1}{2}\gamma^-\gamma^+q$ 
\be
2p^+ \frac{d}{dt}\langle {p}^y \rangle 
&\equiv&
\frac{d}{dt} \left\langle PS\right| \bar{q} \gamma^+
\left(p^y-gA^y\right) q \left| PS \right\rangle
\label{eq:eom1}\\
&=& \frac{1}{\sqrt{2}}
 \frac{d}{dt} \left\langle PS\right| {q}_{+}^\dagger
\left(p^y-gA^y\right) q_{+} \left| PS \right\rangle
\nonumber\\
&=& 2p^+
\left\langle PS\right| \left[\dot{\bar{q}} \gamma^+
\left(p^y-gA^y\right) q + \bar{q}\gamma^+
\left(p^y-gA^y\right) \dot{q}\right]
\left| PS \right\rangle\nonumber\\
& &-  \left\langle PS\right|\bar{q} \gamma^+
g\dot{A}^y q
\left| PS \right\rangle .\nonumber
\ee
Using the QCD equations of motion 
\be
\dot{q} = 
\left(igA^0 + \gamma^0 {\vec \gamma}\cdot {\vec D}\right)q,
\label{eq:eom2}
\ee
where $-iD^\mu = p^\mu-gA^\mu$,
yields
\be
2p^+\frac{d}{dt}\langle {\bf p}^y \rangle &=& 
\left\langle PS\right| \bar{q}\gamma^+ g\left(G^{y0}+G^{yz}\right)
q \left| PS \right\rangle + 
`\left\langle PS\right| \bar{q} \gamma^+ \gamma^- \gamma^i 
D^i D^j q \left| PS \right\rangle'
\nonumber\\
&=& \sqrt{2}
\left\langle PS\right| \bar{q}\gamma^+ gG^{y+}
q \left| PS \right\rangle + 
`\left\langle PS\right| \bar{q} \gamma^+ \gamma^- \gamma^i 
D^i D^j q \left| PS \right\rangle',\label{eq:eom4}
\ee
where $`\left\langle PS\right| \bar{q} \gamma^+ \gamma^- \gamma^i 
D^i D^j q \left| PS \right\rangle'$ stands symbolically for
all terms that involve a product of $\gamma^+\gamma^-$ as well
as a $\gamma^\perp$ and only $\perp$
derivatives $D^i$.

Now it is important to keep in mind that we are not interested in the
average force on the `original' quark 
fields (before the quark is struck by the virtual photon), but
{\it after} absorbing the virtual photon and moving with (nearly)
the speed of light in the $-\hat{z}$ direction.
In this limit, the first term on the r.h.s. of (\ref{eq:eom4})
dominates, as it contains the largest number of `$+$' Lorentz 
indices.
Dropping the other terms yields (\ref{eq:QS3}).

The identification of $2M^2d_2$ with the average transverse force
acting on the active quark in a SIDIS experiment is also
consistent with the 
Qiu Sterman result \cite{QS} for the average transverse
momentum of the ejected quark (also averaged over the momentum
fraction $x$ carried by the active quark)
\be
\langle k_\perp^y\rangle = - \frac{1}{2P^+}
\left\langle P,S \left|\bar{q}(0)\int_0^\infty dx^-G^{+y}(x^+=0,x^-)
\gamma^+q(0) \right|P,S\right\rangle
\label{eq:QS}
\ee
The average transverse 
momentum is obtained by integrating the transverse component
of the color Lorentz force along the trajectory of the active quark
--- which is an almost light-like trajectory along the 
$-\hat{z}$ direction, with $z=-t$. The local twist-3 matrix element
describing the force at time=0 is the first integration point in
the Qiu-Sterman integral (\ref{eq:QS}).

Lattice calculations of the twist-3 matrix element yield 
\cite{latticed2}
\be
d_2^{(u)} = 0.010 \pm 0.012 
\quad \quad \quad \quad
d_2^{(d)} = -0.0056 \pm 0.0050
\ee
renormalized at a scale of $Q^2=5$ GeV$^2$ for the smallest
lattice spacing in Ref. \cite{latticed2}. These numbers are 
also consistent with experimental studies \cite{color}.
Using (\ref{eq:QS3}) these (ancient) lattice results thus
imply
\be
F_{(u)} \approx -100 {\rm MeV/fm}\quad \quad \quad \quad
F_{(d)} \approx 56 {\rm MeV/fm}.
\ee
In the chromodynamic lensing picture, one would have expected
that $F_{(u)}$ and $F_{(d)}$ are of about the same magnitude and with
opposite sign. The same holds in the large $N_C$ limit.
A vanishing Sivers effect for an isoscalar target would be more
consistent with equal and opposite average forces. However, since
the error bars for $d_2$ include only statistical errors, the 
lattice result may not be inconsistent with 
$d_2^{(d)} \sim - d_2^{(u)}$.

The average transverse momentum from the Sivers effect is
obtained by integrating the transverse force to infinity
(along a light-like trajectory) 
$\langle k^y\rangle = \int_0^\infty dt F^y(t)$. This motivates us to
define an `effective range' 
\be
R_{eff} \equiv \frac{\langle k^y\rangle}{F^y(0)}.
\label{eq:Reff}
\ee
Note that $R_{eff}$ depends on how rapidly the correlations fall
off along a light-like direction and it may thus be larger than
the (spacelike) radius of a hadron. 
Of course, unless the functional form of the integrand is known,
$R_{eff}$ cannot really tell us about the range of the FSI,
but if the integrand in (\ref{eq:QS3}) does not oscillate,
(\ref{eq:Reff}) provides a reasonable estimate for the range
over which the integrand in (\ref{eq:QS3}) is significantly
nonzero.

Fits of the Sivers function 
to SIDIS data yield about $|\langle k^y\rangle|\sim
100$ MeV \cite{Mauro}. 
Together with the (average) value for $|d_2|$ from
the lattice this translates into an effective range $R_{eff}$ of 
about 1 fm.
It would be interesting to compare $R_{eff}$ for different quark 
flavors and as a function of $Q^2$, but this requires more
precise values for $d_2$ as well as the Sivers function.

A relation similar to (\ref{eq:QS3}) can be derived for the
$x^2$ moment of the twist-3 scalar PDF $e(x)$. For its
interaction dependent twist-3 part $\bar{e}(x)$ one finds for an
unpolarized target \cite{Yuji} 
\be
4MP^+P^+ e_2 &=& 
g\left\langle p\right|\bar{q}\sigma^{+i}G^{+i}q
\left|P\right\rangle,
\label{eq:odd1}
\ee
where $e_2\equiv \int_0^1 dx x^2\bar{e}(x)$.
The matrix element on the r.h.s. of Eq. (\ref{eq:odd1})
can be related to the average transverse force acting on
a transversely polarized quark in an unpolarized target right after 
being struck by the virtual photon. Indeed, for the
average transverse momentum in the $+\hat{y}$ direction,
for a quark polarized in the $+\hat{x}$ direction, one finds
\be
\langle k^y \rangle = \frac{1}{4P^+}\int_0^\infty dx^-
g\left\langle p\right| \bar{q}(0) \sigma^{+y}G^{+y}(x^-)q(0)\left|
p\right\rangle
\label{eq:odd2}.
\ee
A comparison with Eq. (\ref{eq:odd1}) shows that the average 
transverse force at $t=0$ (right after being struck) on a
quark polarized in the $+\hat{x}$ direction reads
\be
F^y(0) = \frac{1}{2\sqrt{2}p^+} g\left\langle p\right| 
\bar{q} \sigma^{+y}G^{+y}q\left|
p\right\rangle = \frac{1}{\sqrt{2}}MP^+S^x e_2 = \frac{M^2}{2} e_2.
\ee

The impact parameter distribution for quarks polarized in
the $+\hat{x}$ direction \cite{DH} is shifted in the
$+\hat{y}$ direction \cite{latticeBM,hannafious}.
Applying the chromodynamic lensing mechanism implies a force
in the negative $\hat{y}$ direction for these quarks and one
thus expects $e_2<0$ for both $u$ and $d$ quarks. 
Furthermore, since 
$\kappa_\perp>\kappa$, one would expect that in a SIDIS experiment
the $\perp$ force
on a $\perp$ polarized quark in an unpolarized target on average to
be larger than that on unpolarized quarks in a $\perp$ polarized
target, and thus $|e_2| > |d_2|$.

\section{Summary}
The GPD $E^q(x,0,-\Delta_\perp^2)$,
which arises in the `$x$-decomposition' of the contribution
from quark flavor $q$ to the Pauli form factor $F_2^q$
describes the transverse deformation of the unpolarized quark 
distribution in impact parameter space.
That deformation provides a very intuitive mechanism for transverse
SSAs in SIDIS. As a result, the signs of SSAs can be related to the
contribution from quark flavor $q$ to the nucleon anomalous
magnetic moment. 
Quark-gluon correlations appearing in
the $x^2$-moment of the twist-3 part of the polarized parton
distribution $g_2^q(x)$ have a semi-classical interpretation
as the average (enemble average)
transverse force acting on the struck quark
in DIS from a transversely polarized target in the moment after it
has absorbed the virtual photon. Since the direction of that force
can be related to the transverse deformation of PDFs, one can thus
also relate the sign of these quark-gluon correlations to the
contribution from quark flavor $q$ to the nucleon anomalous
magnetic moment. 

Such a correlation between observables that at first appear to 
have little in common also occurs in the chirally
odd sector: the impact parameter space
distribution of quarks with a given transversity
in an unpolarized target can be related to the Boer-Mulders function
describing the left-right asymmetry of quarks with a given 
transversity in SIDIS from an unpolarized target. Furthermore,
semi-classically,
the quark-gluon correlations appearing in the $x^2$-moment
of the twist-3 part of the scalar PDF $e(x)$ describes the
average transverse force acting on a quark with given transversity
immediately after it has absorbed the virtual photon.


{\bf Acknowledgements:}
I would like to thank 
A.Bacchetta, D. Boer, J.P. Chen, Y.Koike, and Z.-E. Mezziani for 
useful discussions. 
This work was supported by the DOE under grant number 
DE-FG03-95ER40965.


\begin{thebibliography}{99}
\bibitem{mb1} M.Burkardt, Phys. Rev. D {\bf 62}, 071503 (2000), 
Erratum-ibid. D {\bf 66}, 119903 (2002); M. Diehl, Eur. Phys. J. 
C {\bf 25}, 223 (2002); J.P. Ralston and B. Pire,
Phys. Rev. D {\bf 66}, 111501 (2002).

\bibitem{IJMPA} M. Burkardt, Int. J. Mod. Phys. A {\bf 18},
173 (2003).

\bibitem{sivers} D.W. Sivers, Phys.\ Rev.\ D {\bf 43}, 261 (1991).

\bibitem{trento} A. Bacchetta et al., Phys.\ Rev.\ D {\bf 70}, 117504
(2004).


\bibitem{mb:SSA} M. Burkardt, Phys.\ Rev. D {\bf 66}, 114005 (2002);
Phys.\ Rev. D {\bf 69}, 057501 (2004).

\bibitem{spectator} S.J. Brodsky, D.S. Hwang, and I. Schmidt,
Nucl. Phys. B {\bf 642}, 344 (2002); M. Burkardt and D.S. Hwang,
Phys. Rev. {\bf D69}, 074032 (2004);
L.P. Gamberg et al., 
Phys. Rev. D {\bf 67}, 071504 (2003); 
D. Boer, S.J. Brodsky, and D.S. Hwang,
Phys. Rev. D {\bf 67}, 054003 (2003); 
A. Bacchetta et al., Phys. Lett {\bf B578}, 109 (2004);       
M. Radici et al., hep-ph/0708.0232; L.P. Gamberg et al. hep-ph/0708.0324; D. Boer et al., Phys.\
Rev.\ D {\bf 67}, 054003 (2003); L.P.~Gamberg et al., Phys.\ Rev.\  D {\bf 67} (2003) 071504.

\bibitem{Metz} S. Meissner et al., Phys.\ Rev.\ D\ {\bf 76}, 034002
(2007)

\bibitem{wigner} A.V. Belitsky, X. Ji, and F. Yuan, 
Phys. Rev. D {\bf 69}, 074014 (2004).

\bibitem{hermes} A. Airapetian et al. ({\sc Hermes} collab.),
Phys.\ Rev.\ Lett.\ {\bf 94}, 012002 (2005).

\bibitem{compass} A. Martin ({\sc Compass} collab.), Czech.
J. Phys. {\bf 56}, F33 (2006).

\bibitem{WW} S. Wandzura and F. Wilczek, Phys. Lett. {\bf 72B},
195 (1977).

\bibitem{Shuryak} E. Shuryak and A.I. Vainshtein, Nucl. Phys. B
{\bf 201}, 141 (1982).

\bibitem{Jaffe} R.L. Jaffe, Comm. Nucl. Part. Phys. {\bf 19}, 239
(1990).

\bibitem{Ji} B.W. Filippone and X. Ji, Adv. Nucl. Phys. {\bf 26}, 1
(2001).

\bibitem{mb:g2} M. Burkardt, arXiv:0811.1206

\bibitem{QS} J. Qiu and G. Sterman, Phys. Rev. Lett. {\bf 67},
2264 (1991).

\bibitem{latticed2} M. G\"ockeler et al., Phys. Rev. D {\bf 72}, 
054507 (2005). 

\bibitem{color} Z.-E. Mezziani et al., hep-ph/0404066.

\bibitem{Mauro} M. Anselmino et al., Eur. Phys. J. A {\bf 39},
89 (2009).


\bibitem{Yuji} Y. Koike and K. Tanaka, Phys. Rev. D {\bf 51}, 6125
(1995).

\bibitem{DH} M. Diehl and P.H\"agler, Eur. Phys. J. {\bf C44}, 87
(2005).

\bibitem{latticeBM} M. G\"ockeler et al. (QCDSF collaboration),
Phys. Rev. Lett. {\bf 98}, 222001 (2007);
Ph. H\"agler et al. (LHPC collaboration), hep-lat/0705.4295.

\bibitem{hannafious} M. Burkardt and B. Hannafious, Phys. Lett.
B {\bf 658}, 130 (2008).

\end{thebibliography}

\end{document}